\documentclass[a4paper]{emulateapj}

\shorttitle{Detection of gas giants with ALMA}
\shortauthors{Mayer et al.}

\begin{document}

\author{Lucio Mayer\altaffilmark{1}, Thomas Peters\altaffilmark{2}, Jaime E. Pineda\altaffilmark{3} and James Wadsley \altaffilmark{4}}

\altaffiltext{1}{Center for Theoretical Astrophysics and Cosmology, Institute for
Computational Science, University of
Zurich, Winterthurerstrasse 190, CH-8057 Z\"{u}rich, Switzerland}
\altaffiltext{2}{Max-Planck-Institut f\"{u}r Astrophysik, Karl-Schwarzschild-Str. 1,
D-85748 Garching, Germany}
\altaffiltext{3}{Max-Planck-Institut f?r extraterrestrische Physik, Giessenbachstrasse 1 85748 Garching, Germany}
\altaffiltext{4}{Department of Physics and Astronomy, McMaster University, Hamilton, ON L8S 4M1, Canada}

\title{Direct detection of  precursors of gas giants formed by gravitational instability with the Atacama Large 
Millimetre/sub-millimetre Array}

\begin{abstract}

Phases of gravitational instability are expected in the early phases of disk evolution, when the disk
mass is still a substantial fraction of the mass of the star.  Disk fragmentation into sub-stellar objects could
occur in the cold exterior part of the disk. Direct detection of massive gaseous clumps on their way to collapse
into gas giant planets would  offer an unprecedented test of the disk instability model. 
Here we use state-of-the-art 3D radiation-hydro simulations of disks undergoing fragmentation into massive gas giants, 
post-processed
with the RADMC-3D ray-tracing code to produce dust continuum emission maps. These are then fed into 
the Common Astronomy Software Applications (CASA) ALMA simulator.  The synthetic maps show that both overdense spiral arms and 
actual clumps at different stages of collapse can be detected with the Atacama Large Millimetre/sub-millimetre Array (ALMA)
in the  full configuration at the distance of the Ophiuchus star forming region (125 pc). The detection of clumps is 
particularly effective at shorter wavelengths  (690 GHz)  combining  two resolutions with multi-scale clean. 
Furthermore, we show that  a flux-based estimate of the mass of a protoplanetary clump can be
from comparable to a factor of 3 higher than 
the gravitationally bound clump mass. The estimated mass depends on the assumed opacity, and on the gas temperature, which
should be set using the input of
radiation-hydro simulations. We conclude that ALMA has the capability to detect ``smoking gun'' systems that are a 
signpost of the disk instability model for gas giant planet formation.

\end{abstract}


\section{Introduction}

Gravitationally unstable (GI) disks are expected in the early phases of star formation, when the disk mass can still
be an appreciable fraction of the stellar mass, during the Class 0/Class I stage.
Whether disk fragmentation is a common outcome of GI and results in long-lived objects that contract to become gas giant
planets (Mayer et al. 2004; Boss 2005), or even lower mass planets via tidal mass loss (Boley et al. 2010; Galvagni \& Mayer 2014), 
is still debated  (Helled et al. 2014). Yet disk instability offers  a natural
explanation for the massive planets on wide orbits discovered via imaging surveys in the last decade
(e.g. Marois et al. 2008) because the conditions required for disk fragmentation, namely a Toomre instability parameter $Q < 1.4$ 
and short radiative cooling timescales, should be satisfied in the disk at $R > 30$ AU (Durisen et al. 2007; Rafikov 2007; 
Clarke 2009; Boley et al. 2010; Meru \& Bate 2010;2012).
Yet direct evidence that disk fragmentation into planetary-sized objects can take place is still
lacking. In disk instability when protoplanets form from condensations in overdense spiral arms they are massive and extended, spanning
2-6 AU in size  for a typical mass of a few Jupiter masses (Boley et al. 2010). 
The first phase of clump collapse should last  $10^3-10^4$ yr (Galvagni et al. 2012),
after which rapid contraction to Jupiter-like densities should occur owing to H$_2$ dissociation 
(Helled et al. 2006).
The initial slow phase of collapse, in which
the protoplanet is still very extended, should be the easiest to detect due to less stringent angular resolution constraints.
The huge step in sensitivity and angular resolution made
possible with the advent of the ALMA observatory prompted us to consider the possible detection of such early stages of planet
formation by disk instability.

Recently several works have indeed focused on detecting spiral structure in gravitationally unstable disks using ALMA 
(Cossins et al. 2010; Douglas et al. 2013; Dipierro et al. 2014; Evans et al. 2015). Similar studies of marginally unstable
disks exhibiting strong spiral pattern have also been carried out  with near-infrared observations of scattered light (Dong et al. 
2015). These studies have been 
motivated
by the recent discovery of several disks with prominent spiral arms, such as MWC 758 (Benisty et al. 2015) and SAO 206462
(Garufi et al. 2013). They do not focus on the detectability of the extreme outcome of GI, namely fragmentation into gas giant
planets or more massive sub-stellar companions, thus they do not address if ALMA could detect a "smoking gun" signature
of planet formation by GI. 
In addition, spiral arms can also be produced by migrating planets (Zhu et al. 2015; Dong et al. 2015; 
Pohl et al. 2015) , or perturbations by nearby
stellar companions, while extended clumps are a unique feature of disk instability. 

With the exception of Douglas et al. (2013), previous works studying the detectability of spiral structure in GI disks
have employed simulations with simple radiative cooling prescriptions for the disk rather than coupling the hydro solver with radiative
transfer.
This applies also to the only previous study of the detectability of GI clumps, which employed  2D simulations of 
very massive embedded protostellar disks fragmenting predominantly into brown-dwarf sized objects (Vorobyov, Zakhozhay \& 
Dunham 2013). The limited treatment of radiation
can strongly affect the resulting temperature structure in the disk (Durisen et al 2007; Boley et al. 2006; Evans et al. 2015), and hence any inference 
concerning detectability.

Here we report on the first study of the detectability of massive protoplanets formed by disk instability 
with ALMA which employs state-of-the-art 3D radiation-hydro SPH simulations. The latter are used to generate 
ALMA images by means of the  ray-tracing code RADMC-3D \footnote { 
http://www.ita.uni-heidelberg.de/$\sim$dullemond/software/radmc-3d/} combined with the ALMA simulator.

\section{Methods}

\subsection{The simulations}

We perform very high resolution 3D radiation hydrodynamics simulations using the
GASOLINE SPH code (Wadsley et al. 2004) with the implementation of an implicit 
scheme for flux-limited diffusion and photospheric
cooling described and thoroughly tested in Rogers \& Wadsley (2011;2012). 
We solve the mono-frequency radiation-hydro equations using
Rosseland mean and Planck mean opacities by means of a look-up table. We use the opacities
of d'Alessio et al (2001).  Stellar irradiation is included in the computation of the initial equilibrium
of the disk but not in the simulation. We adopt an equation of state with a variable adiabatic index
as a function of temperature which includes the effect of hydrogen ionization and molecular dissociation
(Boley et al. 2007; Galvagni et al. 2012).

We model self-gravitating protoplanetary disks without an embedding envelope.
The disk parameters are very similar to those adopted in Boley et al. (2010).
The central star has a mass of $1.35 M_{\odot}$, comparable to that in the HR8799 
exoplanetary system (Marois et al. 2006).
The disk mass is $0.69 M_{\odot}$ out to a radius of $200$ AU. 
A high disk mass,
comprising a significant fraction of the mass of the host star, should be typical
of Class 0-I disks (Greaves \& Rice 2010; Dunham et al. 2014), although occasionally massive systems are found also in
the Class II/T-Tauri stage (Miotello et al. 2014).

\begin{figure*}
\epsscale{0.8}
\plotone{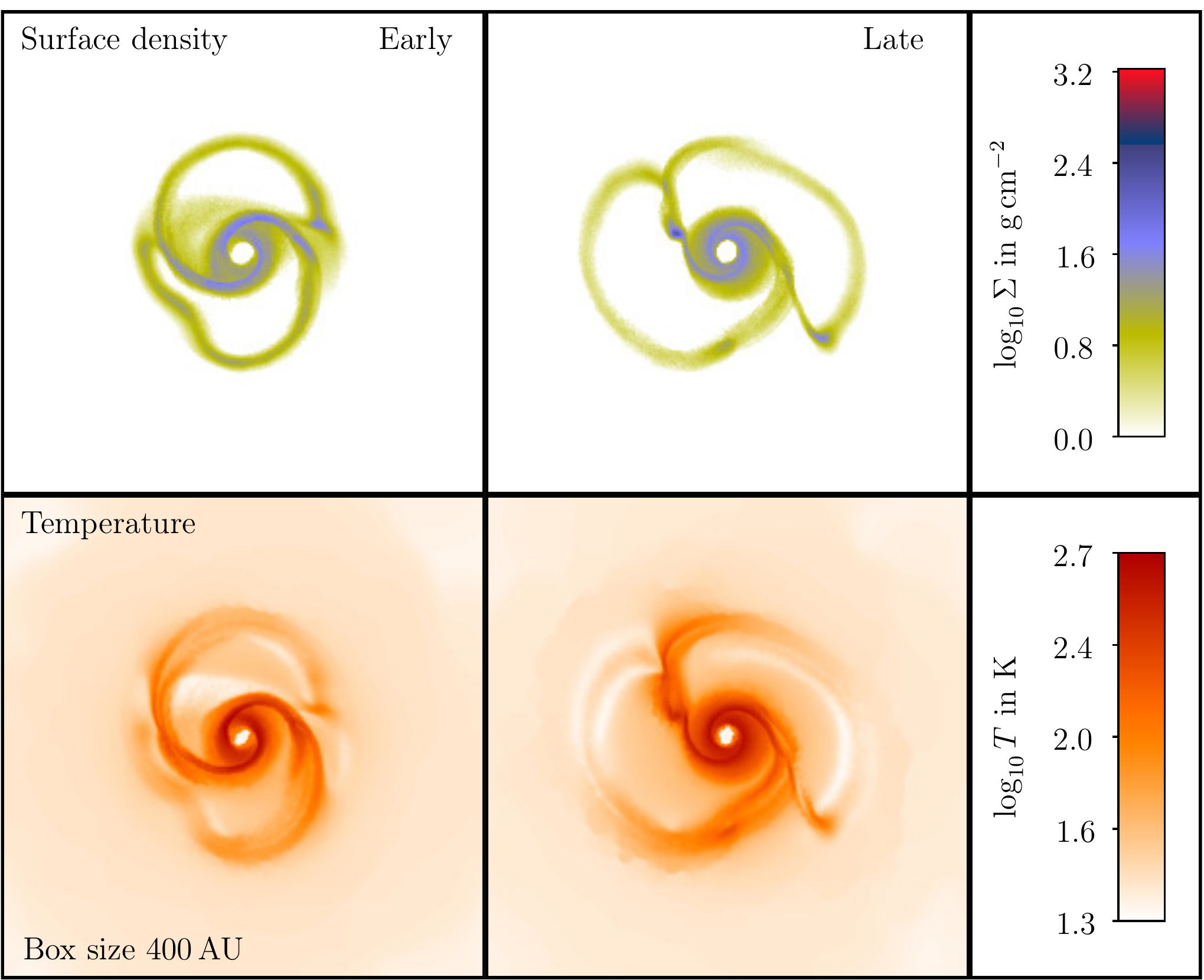}
\caption{Disk surface density (top) and temperature (bottom) at two representative times, immediately
before (early) and one rotation period after fragmentation (late). A slice with thickness equal to one grid cell, namely 0.16 AU,
is shown}.
\end{figure*}

The initial conditions are constructed using an iterative procedure to ensure local balance between pressure, gravity
and centrifugal forces, taking into account the actual gravitational potential of each gas
particle as determined by both the disk and the central star (Rogers \& Wadsley 2011). 
The temperature profile is determined by imposing an initial Toomre Q parameter that 
reaches a minimum of $Q_{min} \sim 1.4$ at $R \sim 60$ AU (see e.g. Durisen et al. 2007).
The simulations comprise 1 million particles in the disk,
with a fixed gravitational softening of $0.16$ AU and a variable SPH smoothing length
which is comparable to the softening at the beginning but can become as small as $0.05$ AU in the highest density regions.
The simulation employed in this paper is part of a set of simulations with different disk masses, stellar
masses and opacities.  Here we focus on one particular simulation, which produces a few massive clumps, one of which is 
gravitationally  bound by the end of the simulation, thereby lending itself naturally for the analysis that we intend to carry out.

\subsection{Post-processing radiative transfer}

We map the SPH data to a homogeneous grid with dimensions $2500 \times 2500 \times
1250$
and a cell size of $0.16\,$AU. We assume a constant gas-to-dust ratio of $100$ and
that gas and dust are collisionally coupled, so that the gas and dust temperatures
are identical. We use the radiative-transfer code
RADMC-3D
to produce synthetic dust emission maps from these data. We follow Dipierro et al. (2014),
who used the opacity law adopted in Cossins et al. (2010):

\begin{equation}
\kappa_\nu = 0.025 \left( \frac{\nu}{10^{12}\,\mathrm{Hz}} \right)
\,\mathrm{cm}^2\,\mathrm{g}^{-1}
\end{equation}
for dust in solar metallicity gas. Note that in this step we adopt a frequency-dependent
opacity law while the simulations simply employed frequency-integrated opacities
to limit the computational burden of the radiative calculation (see previous section).
We produce images of thermal dust emission at four frequencies ($230\,$GHz, $345\,$GHz,
$460\,$GHz and $690\,$GHz) and for five different inclination angles (face-on,
$30^\circ$,
$45^\circ$,  $60^\circ$ and edge-on).

\subsection{ALMA synthetic observations}

We simulate the ALMA full array observations of the dust continuum emission using
tasks \verb+simobserve+ and \verb+simanalyze+ in CASA 4.1.0 (McMullin et al. 2007).
We assume the disk is at the distance of the Ophiuchus star forming
region, 125 pc.
We simulate the continuum ALMA observations for a 10-minutes on-source time, with a
2\,GHz bandwidth, and using 5 different array configurations (alma.out01, alma.out07,
alma.out14, alma.out21, alma.out28). We chose these parameters because they represent
feasible parameters of future snapshot surveys for young disks. The different array
configurations allow for a clear display of the trade-off between sensitivity and angular
resolution. 
Note that the chosen integration time is realistic and at the same time allows to achieve
a good signal-to-noise ratio. For comparison, Dipierro et al. (2014) have considered 
longer integration times (typically 30-120 minutes) in their analysis of spiral structure
detection, while we preferred to be conservative. Clearly for systems located at significantly
larger distances (e.g. 400 pc for Orion) longer integration times will be required in order to approach
the quality of the results presented here.
In addition, we have also combined the synthetic observations of two array configurations:
alma.out14 and alma.out28. The imaging of these combined datasets was done using
standard clean and multi-scale clean. The images created using multi-scale clean show an
improved image fidelity when compared to the standard clean, as we show in the next section.

\section{Results}

The disk quickly develops a prominent spiral pattern that grows in amplitude, quickly leading to overdensities
along the arms. 
The disk fragments into two clumps with masses of several $M_J$ after a few orbits,
at a radius of about 80 AU, where the orbital time is $\sim 2000$ years (Figure 1). A third 
overdensity begins to  form along one of the spiral arms near the end of the simulation.
The simulation is stopped once the first clump that forms , seen at 10 o'clock in Figure 1 
(right panel), 
becomes gravitationally bound and collapses
further, reaching extremely high central densities that render the time-integration prohibitively slow.
The bound mass of the latter clump is $\sim 6.2 M_{J}$ at the last
snapshot, and it has been measured using the SKID group finder with unbinding procedure (see 
\footnote{http://hpcforge.org/projects/skid/}).
The strong spiral pattern, dominated by low-order modes, $m=2-4$, and the masses of the clumps, are fairly typical
of GI-unstable disks undergoing fragmentation (Mayer et al. 2004; 2007; Durisen et al. 
2007; Boley et al. 2010). While the clump has a mass
at the high end of the mass distribution of extrasolar gas giants, we note that small-scale
simulations with much higher resolution, capable of following the collapse of clumps to near-planetary densities, have found that 
the planetary mass resulting at the end of the collapse is at least a factor of 2 lower since a significant fraction
of the mass resides in an extended, loosely bound circumplanetary disk which can be easily stripped by stellar tides
as the protoplanet migrates inward (Galvagni et al. 2012; Galvagni \& Mayer 2014; Malik et al. 2015). 

Our most important result is that massive clumps formed by GI are detectable. Figure 2 shows 
a comparison of the resulting ALMA images for the face-on disk projection. In general
the higher frequency channels, 460 GHz and 690 GHz, are those that best capture the
actual substructure in the disk and its density contrast, separating correctly the clumps, even the more
diffuse ones, from the spiral arms.  At lower frequency the noisy map renders the interpretation
more uncertain, making it difficult to single out even the bound clump. Note that in all 
these images multi-scale clean has been adopted. Its adoption as well as the combination of both
high and mid resolution configuration are crucial, as shown by the comparison in Figure 3. Interestingly,
Figure 3 shows that high resolution alone produces severe artifacts that prevent any identification
of clumps or spiral structure.
Finally Figure 4 shows the comparison of images
obtained for different inclination angles for our best configuration and frequency band.
It is clear that
substructure and its relative contrast can be identified for a range of inclinations.

\begin{figure*}
\epsscale{1.0}
\plotone{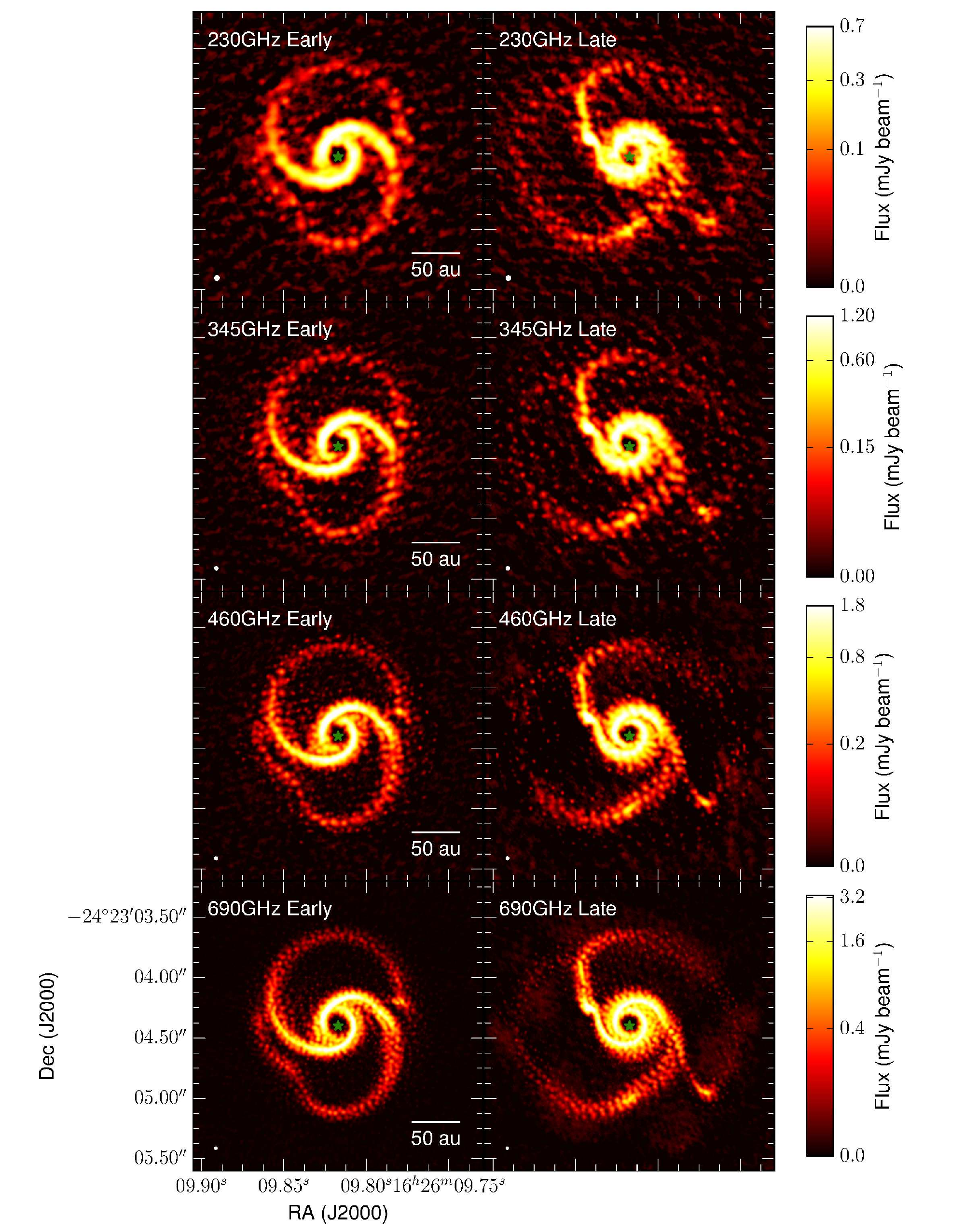}
\caption{Comparison of observed dust continuum emission using two ALMA configurations (C34-14 and C34-28) and multiscale clean. Left and 
right column
present the emission in the Early and Late stages of the disk evolution. From top to bottom, each panels shows the observation at different 
frequency and covering the main bands
available with ALMA. The colorscale is the same for images at the same frequency using an arcsinh stretch, the color bar is shown at the 
right-hand edge. Beam size is shown at the
bottom left corner.}.
\end{figure*}

A key information that one would like to extract from the ALMA images is the mass of the clumps. This is
for two reasons. First, the inferred masses can be used to 
support or refute the hypothesis that the clumps are candidate gas giants rather than e.g. brown dwarf-sized 
objects or simply false detections. Second, by combining the information on mass and size in 
the ALMA images the comparison with simulations can allow to assess if clumps are 
gravitationally bound objects rather than transient over-densities. We thus measured fluxes on the 
maps shown in Figure 2 with an aperture of radius $0.06\arcsec$, corresponding to 7.5 AU at the distance
of Ophiucus, which visually well identifies the (bound) clump in Figure 1. 
This is about a factor of 2 larger than the radius of the bound clump estimated with SKID so it should yield an upper limit on its
mass. Assuming optically-thin emission, a dust-to-gas ratio of 0.01, a temperature of 300 K, and
the Cossins (2010) opacities, we estimate a mass (in increasing frequency) of
18.3, 17.7, 17.1, and 16.5 $M_J$ for this clump.
Using another popular choice of the opacity law,
$\kappa _{\nu}=0.1 (\nu/1.2\,THz) cm^2$/g (Hildebrand, 1983), the estimated masses would be a factor of 3.3 lower,
hence nearly identical to the actual bound mass of the clump.

Mass estimates are extremely sensitive to the assumed gas temperature. If we adopt 30K, 
which is close to the background disk temperature rather than to the temperature of the gas in the
clump region, the inferred mass is largely overestimated, in the range 70-90 $M_J$ even for the
lowest opacities. This would lead to the erroneous conclusion that the clump is a brown dwarf rather than a gas giant.
Therefore the simulations are  instrumental in yielding a good guess on important parameters such as local temperature.
The temperature is however well constrained as spiral shocks in massive gravitationally unstable disks
yield temperature of order 200-400 K quite irrespective of the details of the disk model, hydro code
and radiation solver adopted (see e.g. Mayer et al. 2007; Podolak, Mayer \& Quinn et al. 2011; Boley et al. 2006; Rogers
\& Wadsley 2011). Since spiral arms are the sites of clump formation this is the temperature expected for
clumps soon after their formation. As a gravitationally bound clump collapses further its core temperature eventually approaches
the $H_2$ dissociation temperature of 2000 K eventually, but this occurs on scales of a few Jupiter radii (Helled et al.
2014) that are not resolved with ALMA. Before that happens, however, high resolution simulations of clump
collapse show that the mean temperature rapidly increases to 500-600 K (Galvagni et al. 2012).
Using a temperature of 500 K would yield masses of
$3-10 M_J$ depending on opacity choice, hence very close to the actual bound mass.

Finally, we verified that the inferred mass estimates, for the apparent size of $~ \sim 7.5$ AU
 and for the reference temperature of 300 K, automatically yield that the clump is 
virialized (assuming the scalar virial theorem and spherical symmetry), hence bound. 
The second clump at $\sim 5$ o'clock in Figure 1 (right panel) is not
bound according to  our SKID analysis, and less so is the over-density at 6 o'clock. Note that these
other structures, while weaker in contrast, do show up quite well in the ALMA images at the 
highest frequencies (690 GHz). They appear as extended as they are in the actual simulation at 690 
GHz (and marginally at 480 GHz), while at lower frequencies they are less clear and blend with 
spiral arms, making it difficult to determine their presence as physical substructure in the disk 
(especially for the over-density at 6 o'clock). Note that marginally bound, transient 
over-densities that are easily dissolved by  shear are a recurrent feature of GI disks, hence 
recovering their presence is almost as important as
being able to identify a single bound clump as it is never observed in simulations that 
a fragmenting disk produces only a single clump (Mayer et al. 2004 ; Meru 2015).
Applying our flux-based
mass estimates across the different frequency bands we obtain masses also in the gas
giant planet range for such two overdensities, varying in the range $1-4 M_J$,
depending on the assumed opacity.

\begin{figure}
\epsscale{1.2}
\plotone{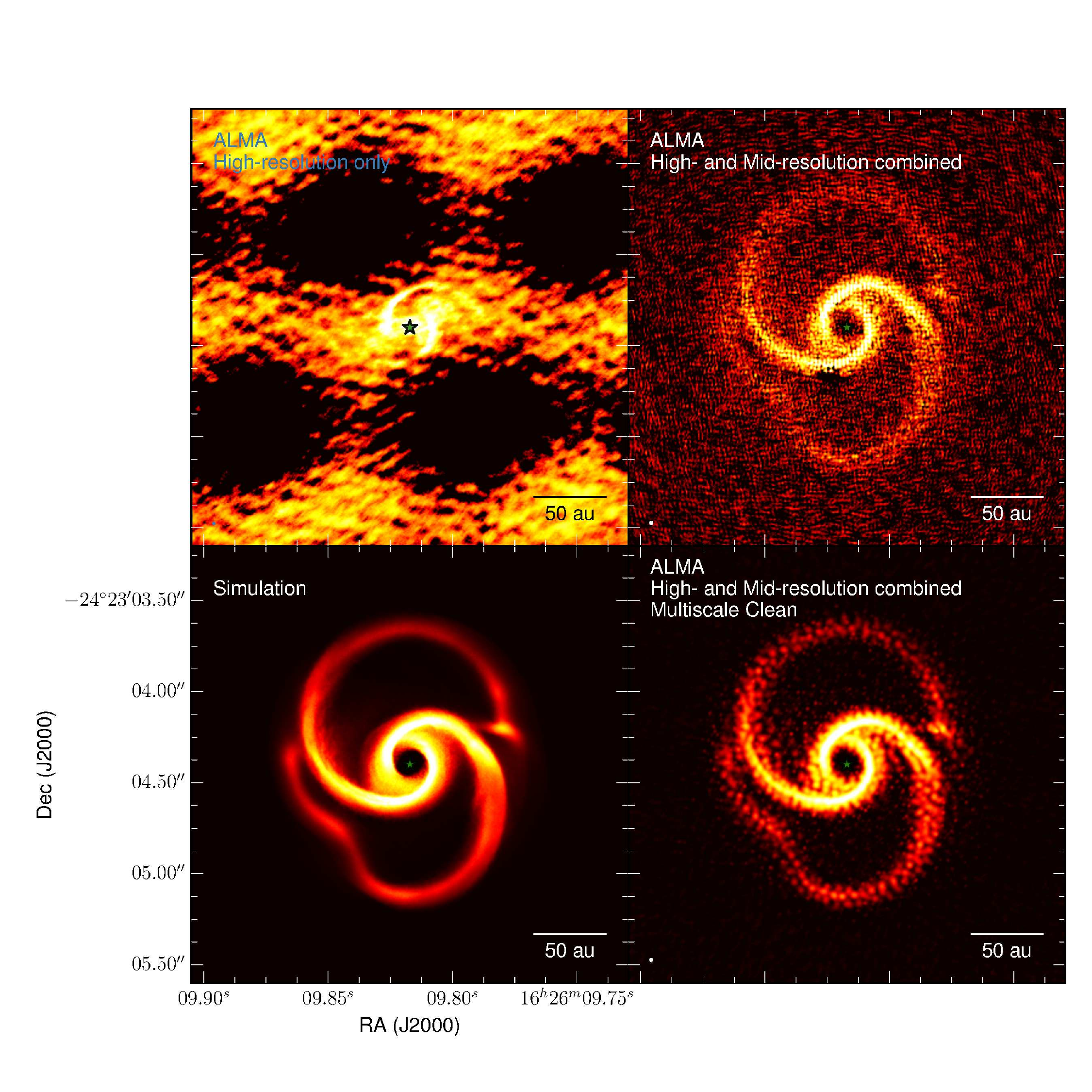}
\caption{Comparison of the simulated disk with three mock observations from ALMA at 690GHz.
Bottom left: dust continuum emission at 690GHz obtained from the radiative transfer of the numerical simulation, this is used as input for 
the ALMA simulated observations.
Top left: ALMA simulated image of the highest angular resolution possible (C34-28) using standard clean. Notice the severe imaging artifacts 
due to the limited uv-coverage, where the
disk emission is more extended.
Top right: ALMA simulated imaged of the combined high- and medium-angular resolution configurations (C34-14 and C34-28) using standard clean.
Bottom right: ALMA simulated imaged of the combined high- and medium-angular resolution configurations (C34-14 and C34-28) using multiscale 
clean.
In all panels the color stretch is between 0 and the peak of the image using an arcsinh stretch. Beam size is shown in the bottom left 
corner.}
\end{figure}


\section{Discussion and Conclusions}

We have reported a proof-of-concept study which combines high-resolution radiation hydro simulations of GI disks with synthetic observations. 
Our study shows that ALMA can detect GI clumps on the scale of gas giants in the early stages of their collapse.  This 
finding extends the results of Dipierro et al. (2014), who showed how even fairly complex spiral structure can be detected by ALMA for a variety of 
frequencies.
While the possibility of clump detection by ALMA was already suggested by Vorobyov et al. (2013), we note that the 2D simulations
in their work considered extremely massive disks, arising soon after the collapse of the molecular cloud core, that 
were almost 
entirely shattered by fragmentation into very massive objects on the scale of brown dwarfs. Such a violent disk instability 
phase, if it ever happens, would last only a couple of rotations, after which the disk itself would disappear.
It would then be hardly observable, and it would not
lead to long-lasting planetary-sized objects (Helled et al. 2014). Furthermore, their synthetic ALMA maps were obtained
from SED modeling rather than by means of a ray-tracing radiative transfer calculation as we do here.

Based on our results detection is possible not only for
very dense, gravitationally bound clumps, which yield the highest density contrast, but also for more loosely bound overdensities
which, while transient, are expected during a GI phase.
High-frequency, combination 
of two (or more) ALMA configurations, and imaging with multi-scale clean provide the optimal setup to capture 
very closely the substructure in the disk at large radii, where the optical depth is relatively low. 
This leads even to fairly accurate mass estimates for clumps, once these can be clearly identified and provided
that a sensible temperature is assumed. Radiation-hydro simulations are crucial in constraining the temperature.

Note that
we find little dependence of the ability to detect clumps and spiral structure on the inclination angle, which is at variance
with Dipierro et al. (2014). However, in the latter work inclination was affecting the simulated ALMA maps when a high-order, tightly
wound spiral pattern was present. In our case the disk, being very massive, develops  only global, large-scale, large pitch
angle $m=2-4$ modes (see also Dong et al. 2015), whose morphology is inherently less affected by inclination, as found also
by Douglas et al. (2013).

Our simulations have some limitations that could have an impact on the ALMA mocks.
We do not include a gaseous envelope, which should still embed the disk in Class 0-I phases, and be the source
of still significant accretion. In principle
protostellar collapse simulations in realistic turbulent cores should be considered.
Note that the envelope could have a non-trivial temperature distribution, on average colder than
the spiral shocks and clumps (with temperatures of a few tens of K), but with hot spots where accretion shocks hit the disk. Since
accretion is filamentary and patchy in a turbulent core (e.g. Hayfield et al. 2011) inhomogeneities in density and
temperature might arise in the outer disk that could render clump identification more difficult. However, using
mocks for both continuum and molecular line emission/absorption, Douglas et al. (2013) have shown that a strong spiral pattern 
should be detectable with ALMA even in embedded disks.

Another important caveat is that here we assumed dust and gas to be distributed in the same
way, both in the simulations and in the post-processing step with RADMC-3D. Recent observations of putative
young protoplanets in HD100546 provide a striking example of how different the distributions of dust and
gas can be (Pineda et al. 2014;  Walsh et al. 2014). In particular, in GI unstable disks dust
would tend to concentrate in spiral arms due to ensuing negative pressure gradients towards the overdensity peaks 
(Rice et al. 2005), and would do even more so after dense clumps have formed (Boley \& Durisen 2010). Therefore 
opacities could be  significantly higher at clump sites, perhaps by an order of magnitude, and also to some extent in spiral arms, 
relative to the background flow. In this case lower frequency observations may have to be considered. The
resulting ALMA mocks will have to be investigated in detail by future work.

\begin{figure}
\epsscale{1.4}
\plotone{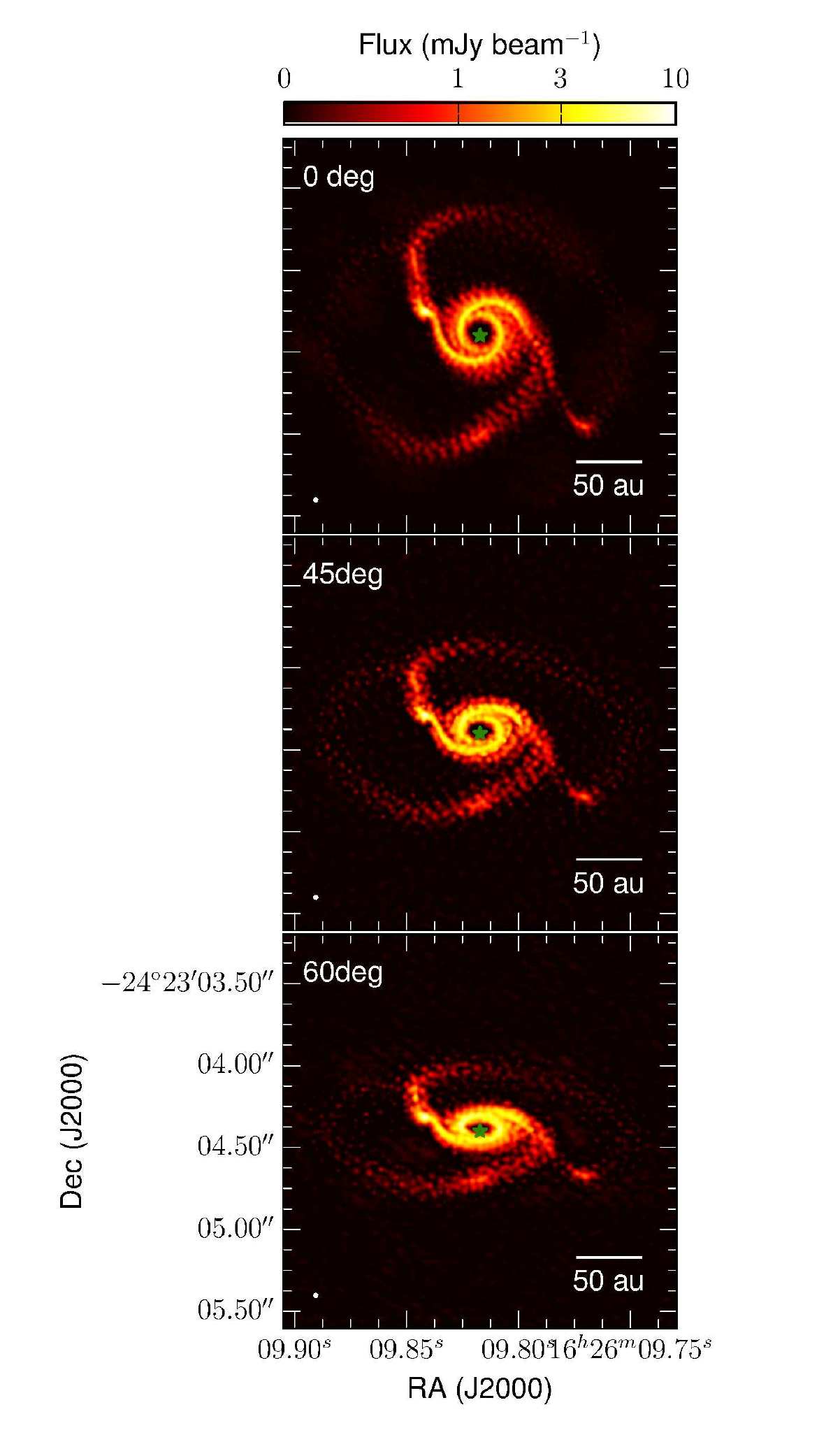}
\caption{Comparison of simulated observations of the dust continuum emission at 690GHz of the disk in a Late stage as observed by ALMA using 
two array configurations (C34-14 and
C34-28) and imaged using multiscale clean. The inclination angles shown are 0 (face-on), 45 and 60 
degrees, and listed in the top left 
corners. The beam size is shown in the bottom
left corner.}
\end{figure}

\bigskip

\acknowledgements

\smallskip

The authors thanh Patrick Rogers for deploying the new disk initial condition generator
used, Marina Galvagni for running the GASOLINE simulations used in this paper, and Joachim Stadel
for improving the TIPGRID code employed to map particle datasets onto grids.
We thank Ravit Helled, Aaron Boley and Farzana Meru for useful comments during preparation of the 
final manuscript.
L.M. thanks the Munich Institute for Astro and Particle Physics (MIAPP)
for hospitality during a crucial phase of this work during summer 2015. 
J.E.P. was supported
by the SINERGIA grant "STARFORM" of the Swiss National Science Foundation during the
early stages of this work, which also enabled the collaboration with L.M and T.P.
J.E.P. acknowledges the financial support of the European Research Council (ERC; project PALs 320620).
T.P. acknowledges support  by a "Forschungskredit" grant of the
University of Z\"urich and by the DFG Priority Program 1573 {\em
Physics of the Interstellar Medium}.

{}

\end{document}